\documentclass[preprint,showpacs,preprintnumbers,amsmath,amssymb,showkeys]{revtex4}

\usepackage{graphicx}
\usepackage{dcolumn}
\usepackage{bm}
\usepackage{braket}
\usepackage{enumitem}
\usepackage{amsmath}
\usepackage{amssymb}
\usepackage{amsthm}
\usepackage{enumitem}
\usepackage[utf8]{inputenc}
\usepackage{amsmath}
\usepackage{amssymb}
\usepackage{amsthm}

\begin{document}

\noindent

\preprint{}

\title{Quantum coherence as asymmetry from complex weak values}

\author{Agung Budiyono$^{1,2,4}$}
\email{agungbymlati@gmail.com}
\author{Mohammad K. Agusta$^{1,2}$}
\author{Bagus E. B. Nurhandoko$^{3}$}
\author{Hermawan K. Dipojono$^{1,2}$}

\affiliation{$^1$Research Center for Nanoscience and Nanotechnology, Bandung Institute of Technology, Bandung, 40132, Indonesia}
\affiliation{$^2$Department of Engineering Physics, Bandung Institute of Technology, Bandung, 40132, Indonesia} 
\affiliation{$^3$Department of Physics, Bandung Institute of Technology, Bandung, 40132, Indonesia} 
\affiliation{$^4$Research Center for Quantum Physics, National Research and Innovation Agency, South Tangerang 15314, Indonesia} 

\date{\today}

\begin{abstract} 

Quantum coherence as an asymmetry relative to a translation group generated by a Hermitian operator, is a necessary resource for the quantum parameter estimation. On the other hand, the sensitivity of the parameter estimation is known to be related to the imaginary part of the weak value of the Hermitian operator generating the unitary imprinting of the parameter being estimated. This naturally suggests a question if one can use the imaginary part of the weak value to characterize the coherence as asymmetry. In this work, we show that the average absolute imaginary part of the weak value of the generator of the translation group, maximized over all possible projective measurement bases, can be used to quantify the coherence as asymmetry relative to the translation group, satisfying certain desirable requirements. We argue that the quantifier of coherence so defined, called TC (translationally-covariant) w-coherence, can be obtained experimentally using a hybrid quantum-classical circuit via the estimation of weak value combined with a classical optimization procedure. We obtain upper bounds of the TC w-coherence in terms of the quantum standard deviation, quantum Fisher information, and the imaginary part of the Kirkwood-Dirac quasiprobability. We further obtain a lower bound and derive a relation between the TC w-coherences relative to two generators of translation group taking a form analogous to the Kennard-Weyl-Robertson uncertainty relation. 

\end{abstract} 

\pacs{03.65.Ta, 03.65.Ca}
\keywords{asymmetry, quantum coherence, weak value, quantum uncertainty, quantum Fisher information, Kirkwood-Dirac quasiprobability}
\maketitle       

\section{Introduction\label{Introduction}}   

Quantum coherence was already recognized in the early days as a fundamental concept which contrasts the quantum and classical worlds. Despite this, it received a rigorous quantitative study only recently. Motivated by the observation that coherence, like entanglement, is a key ingredient in various schemes of quantum technology, many researchers have applied the rigorous mathematical framework of quantum resource theory \cite{Horodecki resource theory} to characterize, quantify and manipulate the quantum coherence. Two approaches to quantum resource theory of coherence have been pursued in the literatures \cite{Marvian - Spekkens speakable and unspeakable coherence}. The first approach defines coherence with respect to an incoherent reference basis wherein its encoding is independent of which elements of the basis appearing in the superposition \cite{Aberg quantifying of superposition,Levi quantum coherence measure,Baumgratz quantum coherence measure,Winter operational resource theory of coherence,Streltsov review,Chitambar physically consistent resource theory of coherence}. This notion of coherence, termed speakable coherence \cite{Marvian - Spekkens speakable and unspeakable coherence}, is a resource in quantum computation \cite{Hillery coherence in decision problems,Matera coherence in quantum algorithm,Ma coherence in quantum algorithm}, quantum cryptography \cite{Ma coherence in key distribution} and quantum random number generator \cite{Ma coherence in random number generator}. The second approach defines coherence as an asymmetry relative to a translation group. In this approach, the encoding of the coherence depends on which elements of the incoherent basis appearing in the superposition, and called unspeakable coherence \cite{Marvian coherence measure quantum speed limit,Marvian application of coherence as asymmetry for aligning reference frame,Marvian coherence as asymmetry 0,Marvian coherence as asymmetry,Marvian - Spekkens speakable and unspeakable coherence,Girolami quantum coherence measure}. Coherence as asymmetry is also called TC (translationally-covariant) coherence since it must not be increasing under translationally-covariant operations \cite{Marvian - Spekkens speakable and unspeakable coherence}. It plays crucial roles in quantum frame alignment \cite{Marvian application of coherence as asymmetry for aligning reference frame}, quantum metrology \cite{Marvian coherence as asymmetry 0,Marvian coherence as asymmetry,Piani robustness of asymmetry}, quantum speed limit \cite{Marvian coherence measure quantum speed limit,Mondal asymmetry and speed limit}, and quantum thermodynamics \cite{Aberg coherence in quantum thermodynamics,Lostaglio coherence in quantum thermodynamics 1,Lostaglio coherence in quantum thermodynamics 2,Cwiklinski coherence in quantum thermodynamics,Yang coherence in quantum thermodynamics}. This latter approach to coherence is the case of interest in the present article. 

Coherence as asymmetry relative to a translation group generated by a Hermitian operator arises naturally in the archetypal scheme of quantum metrology based on quantum parameter estimation \cite{Giovannetti quantum estimation review}. In this scheme, an unknown parameter $\theta\in\mathbb{R}$ is first imprinted to the quantum state $\varrho_{\theta}$ of a probe via a unitary translation generated by a Hermitian operator $K$. The parameter $\theta$ is then estimated from the statistics of the outcomes of some measurement over the quantum state $\varrho_{\theta}$. On the other hand, the sensitivity of such parameter estimation is known to be formally related to the imaginary part of the weak value  \cite{Aharonov weak value,Aharonov-Daniel book,Wiseman weak value,Tamir weak value review,Dressel weak value review} of the generator  $K$ of the unitary imprinting, with the preselected quantum state $\varrho_{\theta}$ \cite{Hofmann weak value and parameter sensitivity}. This naturally raises the question if the imaginary part of the weak value of the generator of the translation group, can provide an insightful and useful characterization of coherence as asymmetry relative to the translation group. Such a characterization is desirable since, first, weak value can be obtained in experiment via a number of methods \cite{Aharonov weak value,Wiseman weak value,Dressel weak value review,Lundeen complex weak value,Jozsa complex weak value,Johansen quantum state from successive projective measurement,Johansen weak value from a sequence of strong measurement,Vallone strong measurement to reconstruct quantum wave function,Cohen estimating of weak value with strong measurements,Lostaglio KD quasiprobability and quantum fluctuation,Wagner measuring weak values and KD quasiprobability,Haapasalo generalized weak value}. Second, a relation between weak value and coherence as asymmetry or TC coherence may open a fresh understanding on the roles of the latter in the broad areas of quantum science wherein the concept of weak value has been shown to be useful. These include quantum state tomography \cite{Lundeen direct measurement of wave function,Lundeen measurement of KD distribution,Maccone comparison between direct state measurement and tomography,Haapasalo generalized weak value}, quantum thermodynamics \cite{Lostaglio contextuality in quantum linear response,Allahverdyan TMH as quasiprobability distribution of work,Levy quasiprobability distribution for heat fluctuation in quantum regime}, quantum information scrambling in many body system \cite{Halpern quasiprobability and information scrambling,Alonso KD quasiprobability witnesses quantum scrambling}, the characterization of different forms of quantum fluctuations \cite{Lostaglio KD quasiprobability and quantum fluctuation}, and  quantum foundation \cite{Pusey strange weak value and contextuality,Lostaglio TMH quasiprobability fluctuation theorem contextuality,Kunjwal contextuality of non-real weak value,Levy quasiprobability distribution for heat fluctuation in quantum regime}. 

In this article, we develop a quantitative characterization of the TC coherence, or coherence as asymmetry in a state $\varrho$ relative to a translation group, in terms of the weak values associated with the Hermitian operator $K$ generating the unitary translation. We first introduce a quantity defined as the average absolute imaginary part of the weak value of the generator of the translation group, optimized over all possible projective bases of the Hilbert space. We argue that it can be used to quantify the TC coherence or coherence as asymmetry relative to the translation group, by showing that it satisfies certain plausible requirements suggested in Ref. \cite{Marvian - Spekkens speakable and unspeakable coherence}, including most importantly, monotonicity under translationally-covariant quantum operations. Accordingly, we call the quantity TC w-coherence. We then argue that the TC w-coherence can be obtained in experiment without recoursing to quantum state tomography via the estimation of the weak value combined with a classical optimization in a hybrid quantum-classical scheme suitable for the present NISQ (Noisy Intermediate-Scale Quantum) hardware. These schemes lend themselves to the operational interpretation of the TC w-coherence. We show that the TC w-coherence is upper bounded by the standard quantum deviation of $K$ in $\varrho$, suggesting a statistical interpretation as a genuine quantum uncertainty arising in the measurement of $K$ over $\varrho$. It gives a lower bound to the square root of the quantum Fisher information about a scalar parameter $\theta$ imprinted by the unitary translation generated by $K$. We further show that the TC w-coherence normalized by the spectral radius of the generator $K$ of the translation group is also upper bounded by the maximum total sum of the absolute imaginary part of the KD (Kirkwood-Dirac) quasiprobability \cite{Kirkwood quasiprobability,Dirac quasiprobability,Chaturvedi KD distribution} which has seen a revival of interest recently to study the nonclassicality arising in a wide spectrum of quantum phenomena \cite{Lostaglio KD quasiprobability and quantum fluctuation}. Finally we obtain a lower bound and derive a relation between TC w-coherences relative to two generators of translation which takes a form analogous to the KWR (Kennard-Weyl-Robertson) uncertainty relation. 

The article is organized as follows. In Section \ref{Preliminaries}, we summarize the concepts of quantum coherence as asymmetry and weak value, and motivate their relation using the setting of quantum parameter estimation. We then define our quantifier of coherence as asymmetry, the TC w-coherence, in terms of the imaginary part of the weak value in Section \ref{Quantifier of TC coherence from the imaginary part of weak value}, and prove that it satisfies certain plausible requirements. We proceed in Section \ref{Observation of TC w-coherence via the observation of weak value} to briefly sketch and discuss the experimental scheme to obtain the TC w-coherence via variational quantum circuits. In Section \ref{Calculation for a single qubit} we give an example of a concrete calculation of TC w-coherence of a single qubit. We then discuss the relation between the TC w-coherence and the quantum variance in Section \ref{TC w-coherence and quantum standard deviation}, with the quantum Fisher information in Section \ref{TC w-coherence and quantum Fisher information}, and with the nonreal values of the KD quasiprobability in Section \ref{TC w-coherence, KD quasiprobability and KWR lower bound}. Finally, in Section \ref{Uncertainty relation for TC w-coherences} we obtain a lower bound for the TC w-coherence, and derive a relation for the TC w-coherences relative to two generators of translation group which takes a form similar to the Kennard-Weyl-Robertson uncertainty relation. The article ends with conclusion and remarks in Section \ref{Conclusion and Remarks}. 

\section{Preliminaries \label{Preliminaries}}  

\subsection{Coherence as translational asymmetry}

To give a physical motivation for coherence as asymmetry and its characterization and  quantification based on complex weak value discussed in the next section, we briefly discuss the following protocol for quantum parameter estimation \cite{Giovannetti quantum estimation review,Holevo book on quantum statistics}. In this article, we shall only discuss quantum system with finite-dimensional Hilbert space. Consider a setting whereby an unknown parameter $\theta\in\mathbb{R}$, whose value we wish to estimate, is imprinted to the quantum state $\varrho_{\theta}\in\mathcal{H}$ of a probe on the Hilbert space $\mathcal{H}$ via a unitary translation generated by a Hermitian operator $K$ conjugate to $\theta$, i.e., 
\begin{eqnarray}
\varrho_{\theta}=U_{K,\theta}\varrho U_{K,\theta}^{\dagger},\hspace{2mm}{\rm with}\hspace{2mm} U_{K,\theta}=e^{-iK\theta},
\label{unitary translation}
\end{eqnarray}
where $\varrho$  is the initial state of the probe. The value of $\theta$ is then inferred from the statistics of the outcomes of some measurement over $\varrho_{\theta}$. Note that the class of unitaries $U_{K,\theta}$, $\theta\in\mathbb{R}$, composes a (representation of) translation group which defines a symmetry. Namely, such a class of translation unitaries single out a set of symmetric states, i.e., states that are left invariant under the translation generated by $K$: $U_{K,\theta}\varrho_{\rm TI}U_{K,\theta}^{\dagger}=\varrho_{\rm TI}$ for all $\theta\in\mathbb{R}$. A symmetric state must therefore commute with the generator of the translation, $[\varrho_{\rm TI},K]=0$, so that they must be jointly diagonalizable. This means that a symmetric state must be a statistical mixture of the elements of the eigenbasis $\{\ket{k}\}$ of $K$, i.e., it is incoherent with respect to the reference basis $\{\ket{k}\}$. \\
{\bf Definition 1}. A state is symmetric relative to the translation group generated by $K$, and thus incoherent with respect to the eigenbasis of $\{\ket{k}\}$ of $K$, if and only if it can be expressed as follows:  
\begin{eqnarray}
\varrho_{\rm TI}:=\sum_k p_k\ket{k}\bra{k}, 
\label{translationally symmetric state}
\end{eqnarray}
where $p_k\ge 0$, $\sum_kp_k=1$.   

It is then clear in the above protocol that in order for the state $\varrho_{\theta}$ to be parameterized by (or, to depend on) the unknown parameter $\theta$ enabling its estimation via a measurement over $\varrho_{\theta}$, the initial state $\varrho$  of the probe must break the translational symmetry, i.e., it cannot be written in the form of (\ref{translationally symmetric state}). A state that is asymmetric relative to the translation group generated by $K$ is a superposition of the eigenbasis $\{\ket{k}\}$ of $K$. Any state that is not translationally invariant, i.e., asymmetric relative to the translation group generated by $K$, must therefore contain some form of coherence relative to the eigenbasis $\{\ket{k}\}$ of $K$. The coherence captures the failure of commutativity between the state and the generator of the translation group: $[K,\varrho]\neq 0$. Following Ref. \cite{Marvian - Spekkens speakable and unspeakable coherence}, we refer to such notion of coherence as TC (translationally-covariant) coherence. Hence, coherence as translational asymmetry is a necessary resource in quantum parameter estimation.

Let us discuss the sensitivity of the above scheme of estimation of $\theta$. To do this, we must make some measurement over $\varrho_{\theta}$. Suppose that the measurement outcomes is denoted by $\{x\}$ with the probability ${\rm Pr}(x|\varrho_{\theta})$. The (classical) Fisher information about $\theta$ contained in ${\rm Pr}(x|\varrho_{\theta})$ is then defined as   
\begin{eqnarray}
J_{\theta}(\{{\rm Pr}(x|\varrho_{\theta})\}):=\sum_x\Big(\frac{\partial_\theta{\rm Pr}(x|\varrho_{\theta})}{{\rm Pr}(x|\varrho_{\theta})}\Big)^2{\rm Pr}(x|\varrho_{\theta}).  
\label{classical Fisher information}
\end{eqnarray}
It quantifies the sensitivity of the estimation of $\theta$ based on $x$ sampled from ${\rm Pr}(x|\varrho_{\theta})$. Indeed, if the estimate $\hat{\theta}(x)$ of $\theta$ is locally unbiased, i.e., $\sum_x\hat{\theta}(x){\rm Pr}(x|\varrho_{\theta})=\theta$, the mean squared error of the estimation is lower bounded by the Fisher information via the Cram\'er-Rao inequality as 
\begin{eqnarray}
\Delta^2_{\hat{\theta}}(\varrho_{\theta})\ge\frac{1}{\nu J_{\theta}(\{{\rm Pr}(x|\varrho_{\theta})\})},
\label{classical Cramer-Rao inequality}
\end{eqnarray} 
where $\nu$ is the number of the probing. We stress that when the initial state $\varrho$  of the probe is symmetric relative to the translation group generated by $K$ so that $\varrho_{\theta}=\varrho$, then ${\rm Pr}(x|\varrho_{\theta})={\rm Pr}(x|\varrho)$ is independent of $\theta$. This implies $J_{\theta}(\{{\rm Pr}(x|\varrho_{\theta})\})=0$ per definition, so that $\Delta^2_{\hat{\theta}}(\varrho_{\theta})$ is infinite. Hence, translational asymmetry is indeed a necessary resource for the parameter estimation via  a nonvanishing Fisher information $J_{\theta}(\{{\rm Pr}(x|\varrho_{\theta})\})$.  

\subsection{Nonreal weak value as signature of coherence as asymmetry}

Let us assume that the probability distribution ${\rm Pr}(x|\varrho_{\theta})$ used to compute the classical Fisher information in Eq. (\ref{classical Fisher information}) is obtained via a measurement described by a complete set of projectors, i.e., projection-valued measure: $\{\Pi_x\}$, $\Pi_x:=\ket{x}\bra{x}$, $\sum_x\Pi_x=\mathbb{I}$, where each $\Pi_x$ is assumed to be rank-one, and $\mathbb{I}$ is the identity operator on the Hilbert space. According to the Born's rule, we thus have ${\rm Pr}(x|\varrho_{\theta})={\rm Tr}(\Pi_x\varrho_{\theta})$. One may consider more general measurement described by POVM (positive-operator-valued measure). However, for our purpose to motivate the construction of coherence quantifier later, at the present moment we consider measurements that are represented by projection-valued measure. We show that the classical Fisher information is formally related to a complex-valued quantity which can be obtained in experiment termed  the weak-value. \\
{\bf Definition 2}. Given an arbitrary preselected state $\varrho$ and a projector $\Pi_x$, the weak value associated with a Hermitian operator $K$ is defined as \cite{Aharonov weak value,Wiseman weak value}: 
\begin{eqnarray}
K^{\rm w}(\Pi_x|\varrho):=\frac{{\rm Tr}(\Pi_xK\varrho)}{{\rm Tr}(\Pi_x\varrho)}. 
\label{complex weak value}
\end{eqnarray} 

Now, from the unitary translation of Eq. (\ref{unitary translation}), we have $\partial_{\theta}\varrho_{\theta}=-i[K,\varrho_{\theta}]$. Noting this, the imaginary part of the weak value $K^{\rm w}(\Pi_x|\varrho_{\theta})$ can thus be expressed as 
\begin{eqnarray}
{\rm Im}(K^{\rm w}(\Pi_x|\varrho_{\theta}))=\frac{{\rm Tr}(\Pi_x[K,\varrho_{\theta}])}{2i{\rm Tr}(\Pi_x\varrho_{\theta})}=\frac{1}{2}\frac{\partial_{\theta}{\rm Pr}(x|\varrho_{\theta})}{{\rm Pr}(x|\varrho_{\theta})}. 
\label{imaginary part of weak value and score function}
\end{eqnarray}
Using Eq. (\ref{imaginary part of weak value and score function}), the Fisher information of Eq. (\ref{classical Fisher information}) can therefore be written in terms of the variance of the imaginary part of the weak value of the generator $K$ as
\begin{eqnarray}
J_{\theta}(\{{\rm Pr}(x|\varrho_{\theta})\})=4\sum_x({\rm Im}(K^{\rm w}(\Pi_x|\varrho_{\theta})))^2{\rm Pr}(x|\varrho_{\theta}). 
\end{eqnarray}
Hence, the imaginary part of the weak value $K^{\rm w}(\Pi_x|\varrho_{\theta})$ characterizes the sensitivity of estimation of $\theta$ via the Fisher information \cite{Hofmann weak value and parameter sensitivity}. In particular, a nonvanishing ${\rm Im}(K^{\rm w}(\Pi_x|\varrho_{\theta}))$ is necessary and sufficient for a non-vanishing Fisher information $J_{\theta}(\{{\rm Pr}(x|\varrho_{\theta})\})$. Since a nonvanishing $J_{\theta}(\{{\rm Pr}(x|\varrho_{\theta})\})$ indicates an asymmetry of $\varrho$ relative to the translation group generated by $K$, the above observation naturally suggests a question if one may use ${\rm Im}\{K^{\rm w}(\Pi_x|\varrho_{\theta})\}$ to characterize and quantify the coherence as asymmetry of $\varrho$ relative to the translation group generated by $K$. 

\section{Quantifying TC coherence using complex weak values}

\subsection{TC w-coherence: a quantifier of TC coherence from the imaginary part of weak value\label{Quantifier of TC coherence from the imaginary part of weak value}}

We show in this section that the imaginary part of the weak value of an arbitrary Hermitian operator $K$ can indeed be used to characterize and quantify quantum coherence as asymmetry in a quantum state with respect to the translation group generated by $K$. \\ 
{\bf Definition 3}. Given a general quantum state $\varrho$ on a finite-dimensional Hilbert space and a Hermitian operator $K$ generating a translation group, we define a quantity, called TC w-coherence, which maps the quantum state and the generator to a nonnegative real number as:
\begin{eqnarray}
C_{\rm w}(\varrho;K)&:=&\sup_{\{\ket{x}\}}\sum_x\big|{\rm Im}K^{\rm w}(\Pi_x|\varrho)\big|{\rm Pr}(x|\varrho)\nonumber\\
&=&\sup_{\{\ket{x}\}}\sum_x\big|{\rm Im}\braket{x|K\varrho|x}\big|\nonumber\\
&=&\frac{1}{2}\sup_{\{\ket{x}\}}\sum_x\big|\braket{x|[K,\varrho]|x}\big|. 
\label{TC w-coherence}
\end{eqnarray}\\
Hence, we take the average of the absolute imaginary part of the weak value $K^{\rm w}(\Pi_x|\varrho)$ defined in Eq. (\ref{complex weak value}) over the probability to get the outcome $x$ in a measurement represented by a projection-valued measure $\{\Pi_x\}$, i.e., ${\rm Pr}(x|\varrho)={\rm Tr}(\Pi_x\varrho)$, and maximize over all the projective bases $\{\ket{x}\}$ of the Hilbert space. From the last line of Eq. (\ref{TC w-coherence}), $C_{\rm w}(\varrho;K)$ can be seen to measure the failure of commutativity between the quantum state and the generator of the translation group. 

Next, consider a composite of $N$ subsystems and suppose that the generator $K$ of the translation group is local, i.e., it is the sum of tensor product of local operators. Then, in this case, the defining basis $\{\ket{x}\}$ in Eq. (\ref{TC w-coherence}) is given by the tensor product of the basis of each subsystem, i.e., $\ket{x}=\otimes_{i=1}^N\ket{x_i}$, where $\{\ket{x_i}\}$ is the basis of the subsystem $i$, $i=1,\dots,N$. \\
{\bf Remark}. The TC w-coherence in a multipartite state relative to the translation group with a local generator $K$ is given by:
\begin{eqnarray}
&&C_{\rm w}(\varrho_{1\cdots N};K)\nonumber\\
&:=&\sup_{\{\ket{x_1,\dots,x_N}\}}\sum_{x_1,\dots,x_N}\big|{\rm Im}K^{\rm w}(\Pi_{x_1}\otimes\cdots\otimes\Pi_{x_N}|\varrho_{1,\dots,N})\big|{\rm Pr}(x_1,\dots,x_N|\varrho_{1,\dots,N})\nonumber\\
&=&\sup_{\{\ket{x_1,\dots,x_N}\}}\sum_{x_1,\dots,x_N}\big|{\rm Im}\braket{x_1,\dots,x_N|K\varrho_{1,\cdots,N}|x_1,\dots,x_N}\big|,
\label{TC w-coherence for a composite}
\end{eqnarray}
where $\varrho_{1\cdots N}$ is an $N$-partite state. Note further that while we have motivated the above definition using the setting of the quantum parameter estimation, it applies for general $\varrho$ and $K$. 

We show that $C_{\rm w}(\varrho;K)$ satisfies certain requirements expected for a quantifier of TC quantum coherence or coherence as asymmetry as follows \cite{Marvian - Spekkens speakable and unspeakable coherence,Streltsov review}:\\
{\bf Proposition 1}. {\it Faithfulness}: $C_{\rm w}(\varrho;K)=0$ if and only if the quantum state $\varrho$ is invariant with respect to the translation group generated by $K$, that is, if and only if it is incoherent relative to the eigenbasis $\{\ket{k}\}$ of $K$.\\
{\bf Proof}. First, let us assume that $\varrho$ is incoherent with respect to the eigenbasis $\{\ket{k}\}$ of $K$. $\varrho$ can thus be expressed in the form of (\ref{translationally symmetric state}) so that it commutes with $K$, i.e., $[K,\varrho]=0$. We thus get $C_{\rm w}(\varrho;K)=0$, as per definition of Eq. (\ref{TC w-coherence}). Conversely, suppose that $C_{\rm w}(\varrho;K)=0$. Then, from the definition of Eq. (\ref{TC w-coherence}), we must have $\braket{x|[K,\varrho]|x}=0$ for all the bases $\{\ket{x}\}$ of the Hilbert space. This can only be true if $[K,\varrho]=0$, so that they are simultaneously diagonalizable. Denoting the set of eigenvalues of $\varrho$ as $\{p_k\}$, $p_k\ge 0$, $\sum_kp_k=1$, $\varrho$ must therefore be decomposed as a statistical mixture of the eigenprojectors $\{\ket{k}\bra{k}\}$ of $K$ as in Eq. (\ref{translationally symmetric state}). Hence, $\varrho$ is incoherent relative to $\{\ket{k}\}$, i.e., it is invariant with respect to the translation group generated by $K$. \qed
\\
{\bf Proposition 2}. {\it Convexity}: $C_{\rm w}(\sum_jp_j\varrho_j;K)\le\sum_jp_jC_{\rm w}(\varrho_j;K)$, where $p_j\ge 0$, $\sum_jp_j=1$.\\
{\bf Proof}. This is a straightforward implication of the definition of the TC w-coherence of Eq. (\ref{TC w-coherence}) by virtue of the triangle inequality and the fact that $p_k\ge 0$, i.e., $C_{\rm w}(\sum_jp_j\varrho_j;K)=\sup_{\{\ket{x}\}}\sum_x\big|{\rm Im}\braket{x|K\sum_jp_j\varrho_j|x}\big|\le\sum_jp_j\sup_{\{\ket{x}\}}\sum_x\big|{\rm Im}\braket{x|K\varrho_j|x}\big|=\sum_jp_jC_{\rm w}(\varrho_j;K)$. \qed
\\
{\bf Proposition 3}. {\it Unitary invariance}: $C_{\rm w}(V\varrho V^{\dagger};VKV^{\dagger})=C_{\rm w}(\varrho;K)$ for any unitary transformation $V$.\\
{\bf Proof}. This can be straightforwardly proven as follows:
\begin{eqnarray}
C_{\rm w}(V\varrho V^{\dagger};VKV^{\dagger})&=&\sup_{\{\ket{x}\}}\sum_x\big|{\rm Im}\braket{x|VK V^{\dagger}V\varrho V^{\dagger}|x}\big|\nonumber\\
\label{proof of the unitary covariant property step 3}
&=&\sup_{\{\ket{x'}\}}\sum_{x'}\big|{\rm Im}\braket{x'|K\varrho|x'}\big|\\
\label{proof of the unitary covariant property step 4}
&=&C_{\rm w}(\varrho;K), 
\end{eqnarray}
where we have defined a new basis $\{\ket{x'}\}=\{V^{\dagger}\ket{x}\}$ to get Eq. (\ref{proof of the unitary covariant property step 3}), and Eq. (\ref{proof of the unitary covariant property step 4}) holds since the set of the new bases $\{\ket{x'}\}$ is the same as the set of the old bases $\{\ket{x}\}$ given by all the bases of the same Hilbert space, so that $\sup_{\{\ket{x}\}}(\cdot)=\sup_{\{\ket{x'}\}}(\cdot)$. \qed
\\
{\bf Proposition 4}. {\it Invariance under translationally-covariant unitaries}: $C_{\rm w}(V\varrho V^{\dagger};K)=C_{\rm w}(\varrho;K)$, where the unitary transformation $V$ is said translationally-covariant with respect to the translation group generated by $K$ if the result of its implementation followed by the translation generated by $K$ yields the same effect when the order of the operations is reversed, i.e, $U_{K,\theta} V\varrho V^{\dagger}U_{K,\theta}^{\dagger}=VU_{K,\theta}\varrho U_{K,\theta}^{\dagger}V^{\dagger}$ \cite{Marvian - Spekkens speakable and unspeakable coherence}, where $U_{K,\theta}$ is the translation unitary given in Eq. (\ref{unitary translation}). It maps incoherent state to incoherent state. \\
{\bf Proof}. First, for a unitary $V$ to be translationally-covariant with respect to the group of translation generated by $K$, it must commute with $K$, i.e. $VK=KV$ so that $V^{\dagger}KV=K$ \cite{Marvian - Spekkens speakable and unspeakable coherence}. Noting this, we thus have 
\begin{eqnarray}
&&C_{\rm w}(V\varrho V^{\dagger};K)\nonumber\\
\label{proof: nonincreasing under translationally invariant unitary step 1}
&=&\sup_{\{\ket{x}\}}\sum_x\big|{\rm Im}\braket{x|VV^{\dagger}KV\varrho V^{\dagger}|x}\big|\\
\label{proof: nonincreasing under translationally invariant unitary step 2}
&=&\sup_{\{\ket{x}\}}\sum_{x'}\big|{\rm Im}\braket{x'|K\varrho|x'}\big|\\
\label{proof: nonincreasing under translationally invariant unitary step 3}
&=&C_{\rm w}(\varrho;K),
\end{eqnarray}
where we have inserted the identity $ VV^{\dagger}=\mathbb{I}$ in Eq. (\ref{proof: nonincreasing under translationally invariant unitary step 1}), to get Eq. (\ref{proof: nonincreasing under translationally invariant unitary step 2}) we have defined the new basis via unitary transformation $\{\ket{x'}\}=\{V^{\dagger}\ket{x}\}$, and Eq. (\ref{proof: nonincreasing under translationally invariant unitary step 3}) holds due to the fact that $\sup_{\{\ket{x}\}}(\cdot)=\sup_{\{\ket{x'}\}}(\cdot)$. \qed
\\
{\bf Proposition 5}. {\it Non-increasing under partial trace}: $C_{\rm w}(\varrho_{12};K_1\otimes\mathbb{I}_2)\ge C_{\rm w}(\varrho_1;K_1)$, and equality is obtained for a product state, i.e., $\varrho_{12}=\varrho_1\otimes\varrho_2$. Here, $\varrho_{12}\in\mathcal{H}_{12}$ is a density operator on Hilbert space   $\mathcal{H}_{12}=\mathcal{H}_1\otimes\mathcal{H}_2$ of a bipartite system 1 and 2, $\varrho_1={\rm Tr}_2\{\varrho_{12}\}\in\mathcal{H}_1$, $K_1$ is the generator of translation group on $\mathcal{H}_1$, and $\mathbb{I}_2$ is the identity operator on $\mathcal{H}_2$. \\
{\bf Proof}. This can be shown directly as
\begin{eqnarray}
&&C_{\rm w}(\varrho_{12};K_1\otimes\mathbb{I}_2)\nonumber\\
&=&\sup_{\{\ket{x_1,x_2}\}}\sum_{x_1,x_2}\big|{\rm Im}\braket{x_1,x_2|(K_1\otimes\mathbb{I}_2)\varrho_{12}|x_1,x_2}\big|\nonumber\\
&\ge&\sup_{\{\ket{x_1,x_2}\}}\sum_{x_1}\big|{\rm Im}\big(\sum_{x_2}\braket{x_1,x_2|(K_1\otimes\mathbb{I}_2)\varrho_{12}|x_1,x_2}\big)\big|\nonumber\\
&=&\sup_{\{\ket{x_1}\}}\sum_{x_1}\big|{\rm Im}\braket{x_1|K_1\varrho_1|x_1}\big|\nonumber\\
&=& C_{\rm w}(\varrho_1;K_1), 
\end{eqnarray}
where we have used the definition of TC w-coherence in Eq. (\ref{TC w-coherence for a composite}). One can see that, as intuitively expected, the equality is obtained when there is no correlation what so ever between the two systems, i.e., $\varrho_{12}=\varrho_1\otimes\varrho_2$, by virtue of the fact that $\braket{x_2|\varrho_2|x_2}$ is real and nonnegative for all $x_2$ and $\sum_{x_2}\braket{x_2|\varrho_2|x_2}=1$. This property shows that if two systems are correlated, ignoring one of them should not increase the coherence of the other. \qed
\\
{\bf Proposition 6}. {\it Monotonicity}: Non-increasing under a class of completely positive trace nonincreasing linear maps or quantum operation $\Phi(\varrho_s)$, i.e., $C_{\rm w}(\Phi(\varrho_s);K_s)\le C_{\rm w}(\varrho_s;K_s)$, where $\Phi(\varrho_s)$ admits the following free Stinespring dilation \cite{Marvian - Spekkens speakable and unspeakable coherence}: $\Phi(\varrho_s)={\rm Tr}_a\big((\mathbb{I}_s\otimes E_a)V_{sa}(\varrho_s\otimes\varrho_a)V_{sa}^{\dagger}\big)$. Here, $\varrho_s$ is the quantum state of the (principal) system of interest, $\varrho_a$ is the quantum state of an ancilla (auxiliary system), $V_{sa}$ is a unitary which couples the system and the ancilla, and $0\le E_a\le \mathbb{I}_a$ is an effect (i.e., an element of POVM representing a measurement) on the ancilla. Moreover, by free we mean that $V_{sa}$ is covariant with respect to the translation unitary generated by $K_s\otimes\mathbb{I}_a+\mathbb{I}_s\otimes K_a$, where $K_s$ and $K_a$ are the local generators of the translation associated with the system and the ancilla, respectively, and $\varrho_a$ and $E_a$ are incoherent relative to translation generated by $K_a$ so that $[\varrho_a,K_a]=[E_a,K_a]=0$. We note that, as shown in Ref. \cite{Marvian - Spekkens speakable and unspeakable coherence} (see their Proposition 2), such a map is covariant relative to the group of translation generated by $K_s$, i.e., it satisfies $U_{K_s,\theta}\Phi(\varrho_s)U_{K_s,\theta}^{\dagger}=\Phi(U_{K_s,\theta}\varrho_sU_{K_s,\theta}^{\dagger})$. Moreover, conversely, any such translationally-covariant quantum operation can be implemented using the above free Stinespring dilation. A translationally-covariant quantum operation maps incoherent state onto incoherent state. Hence, it generalizes Proposition (iv) to nonunitary dynamics in open systems. \\
{\bf Proof}. First, we have, from the Stinespring dilation theorem, 
\begin{eqnarray}
&&C_{\rm w}(\Phi(\varrho_s);K_s)\nonumber\\
\label{proof of monotonicity 1 step 2}
&=&\sup_{\{\ket{x_s}\}}\sum_{x_s}\big|{\rm Im}\braket{x_s|K_s{\rm Tr}_a((\mathbb{I}_s\otimes E_a)V_{sa}(\varrho_s\otimes\varrho_a)V_{sa}^{\dagger})|x_s}\big|\\
\label{proof of monotonicity 1 step 3}
&\le&\sup_{\{\ket{x_s}\}}\sum_{x_s,k_a}\big|{\rm Im}\langle x_s,k_a|(K_s\otimes\mathbb{I}_a)(\mathbb{I}_s\otimes E_a)V_{sa}(\varrho_s\otimes\varrho_a)V_{sa}^{\dagger}|x_s,k_a\rangle\big|\\
\label{proof of monotonicity 1 step 3.5}
&=&\sup_{\{\ket{x_s}\}}\sum_{x_s,k_a}\big|e_a{\rm Im}\langle x_s,k_a|(K_s\otimes\mathbb{I}_a)V_{sa}(\varrho_s\otimes\varrho_a)V_{sa}^{\dagger}|x_s,k_a\rangle\big|\\
\label{proof of monotonicity 1 step 4}
&=&\sup_{\{\ket{x_s}\}}\sum_{x_s,k_a}\big|e_a{\rm Im}\langle x_s,k_a|(K_s\otimes\mathbb{I}_a+\mathbb{I}_s\otimes K_a)V_{sa}(\varrho_s\otimes\varrho_a)V_{sa}^{\dagger}|x_s,k_a\rangle\big|\\
\label{proof of monotonicity 1 step 5}
&=&\sup_{\{\ket{x_s}\}}\sum_{x_s,k_a}\big|e_a{\rm Im}\langle x_s,k_a|V_{sa}(K_s\otimes\mathbb{I}_a+\mathbb{I}_s\otimes K_a)(\varrho_s\otimes\varrho_a)V_{sa}^{\dagger}|x_s,k_a\rangle\big|\\
\label{proof of monotonicity 1 step 6}
&=&\sup_{\{\ket{x_s}\}}\sum_{x_s,k_a}\big|e_a{\rm Im}\langle x_s,k_a|V_{sa}(K_s\otimes\mathbb{I}_a)(\varrho_s\otimes\varrho_a)V_{sa}^{\dagger}|x_s,k_a\rangle\big|. 
\label{proof of monotonicity 1}
\end{eqnarray} 
Here, the choice of the basis of the ancilla $\{\ket{k_a}\}$ inserted in Eq. (\ref{proof of monotonicity 1 step 3}) is arbitrary, to get Eq. (\ref{proof of monotonicity 1 step 3.5}) we have chosen $\{\ket{k_a}\}$ which diagonalizes $E_a$, i.e., $E_a\ket{k_a}=e_a\ket{k_a}$ with $e_a\in\mathbb{R}$, to get Eq. (\ref{proof of monotonicity 1 step 4}) we have used the assumption that $[E_a,K_a]=0$ so that $\{\ket{k_a}\}$ also diagonalizes $K_a$ to have: ${\rm Im}\langle x_s,k_a|(\mathbb{I}_s\otimes K_a)V_{sa}(\varrho_s\otimes\varrho_a)V_{sa}^{\dagger}|x_s,k_a\rangle=k_a{\rm Im}\langle x_s,k_a|V_{sa}(\varrho_s\otimes\varrho_a)V_{sa}^{\dagger}|x_s,k_a\rangle=0$ where $K_a\ket{k_a}=k_a\ket{k_a}$ with $k_a\in\mathbb{R}$, to get Eq. (\ref{proof of monotonicity 1 step 5}) we have used the assumption that $V_{sa}$ is covariant with respect to the translation generated by $K_s\otimes\mathbb{I}_a+\mathbb{I}_s\otimes K_a$ so that they must commute, i.e., $(K_s\otimes\mathbb{I}_a+\mathbb{I}_s\otimes K_a)V_{sa}=V_{sa}(K_s\otimes\mathbb{I}_a+\mathbb{I}_s\otimes K_a)$, and to get Eq. (\ref{proof of monotonicity 1 step 6}) we have used the assumption $[K_a,\varrho_a]=0$ so that we have: ${\rm Im}\langle x_s,k_a|V_{sa}(\mathbb{I}_s\otimes K_a)(\varrho_s\otimes\varrho_a)V_{sa}^{\dagger}|x_s,k_a\rangle=\langle x_s,k_a|V_{sa}(\varrho_s\otimes[K_a,\varrho_a])V_{sa}^{\dagger}|x_s,k_a\rangle/2i=0$. 

Next, to further evaluate the right-hand side of Eq. (\ref{proof of monotonicity 1}), we first recall that the interaction unitary $V_{sa}$ commutes with $K_s\otimes\mathbb{I}_a+\mathbb{I}_s\otimes K_a$ having the set of eigenvectors $\{\ket{k_s}\otimes\ket{k_a}\}$, where $\{\ket{k_s}\}$ is the set of orthonormal eigenvectors of $K_s$. This means the interaction unitary can be decomposed as $V_{sa}^{\dagger}=\sum_{k_s,k_a}e^{ig_{k_s,k_a}}\ket{k_s}\bra{k_s}\otimes\ket{k_a}\bra{k_a}$, $g_{k_s,k_a}\in\mathbb{R}$. Using this we have 
\begin{eqnarray}
V_{sa}^{\dagger}\ket{x_s,k_a}&=&\big(\sum_{k_s,k_a'}e^{ig_{k_s,k_a}}\ket{k_s}\bra{k_s}\otimes\ket{k_a'}\bra{k_a'}\big)\ket{x_s}\otimes\ket{k_a}\nonumber\\
&=&\big(\sum_{k_s}e^{ig_{k_s,k_a}}\braket{k_s|x_s}\ket{k_s}\big)\otimes\ket{k_a}=\ket{x_s^a}\otimes\ket{k_a},
\label{the new product basis}
\end{eqnarray}  
where $\ket{x_s^a}:=\sum_{k_s}e^{ig_{k_s,k_a}}\braket{k_s|x_s}\ket{k_s}=U_s^a\ket{x_s}$, with $U_s^a:=\sum_{k_s}e^{ig_{k_s,k_a}}\ket{k_s}\bra{k_s}$ is a unitary transformation. Hence, for each $a$, $\{\ket{x_s^a}\}$ comprises a new basis for the Hilbert space of the system, i.e., it is orthonormal and complete. Using Eq. (\ref{the new product basis}) in Eq. (\ref{proof of monotonicity 1}), we thus finally obtain 
\begin{eqnarray}
&&C_{\rm w}(\Phi(\varrho_s);K_s)\nonumber\\
\label{proof of monotonicity step 0}
&\le&\sup_{\{\ket{x_s^a}\}}\sum_{x_s^a,k_a}\big|e_a{\rm Im}\langle x_s^a,k_a|(K_s\otimes\mathbb{I}_a)(\varrho_s\otimes\varrho_a)|x_s^a,k_a\rangle\big|\\
&=&\sup_{\{\ket{x_s^a}\}}\sum_{x_s^a,k_a}\big|{\rm Im}\langle x_s^a,k_a|(K_s\otimes E_a)(\varrho_s\otimes\varrho_a)|x_s^a,k_a\rangle\big|\nonumber\\
\label{proof of monotonicity step 1}
&=&\sup_{\{\ket{x_s^a}\}}\sum_{x_s^a,k_a}\big|{\rm Im}\langle x_s^a|K_s\varrho_s|x_s^a\rangle\big|\braket{k_a|E_a\varrho_a|k_a}\\
&=&\sum_{k_a}C_{\rm w}(\varrho_s;K_s)\braket{k_a|E_a\varrho_a|k_a}=C_{\rm w}(\varrho_s;K_s){\rm Tr}\{E_a\varrho_a\}\nonumber\\
\label{proof of monotonicity step 3}
&\le&C_{\rm w}(\varrho_s;K_s), 
\end{eqnarray}
where Eq. (\ref{proof of monotonicity step 0}) holds because $\sup_{\{\ket{x_s}\}}(\cdot)=\sup_{\{\ket{x_s^a}\}}(\cdot)$ for each $a$, to get Eq. (\ref{proof of monotonicity step 1}) we have taken into account the fact that $\braket{k_a|E_a\varrho_a|k_a}$ is real and nonnegative for all $k_a$ ($E_a$ and $\varrho_a$ commute and both are positive operators), and the inequality in Eq. (\ref{proof of monotonicity step 3}) holds since ${\rm Tr}(E_a\varrho_a)\le1$ ($E_a$ is an element of POVM describing measurement on the auxiliary system).  \qed 

Hence, $C_{\rm w}(\varrho;K)$ defined in Eq. (\ref{TC w-coherence}) based on the imaginary part of the weak value of $K$ with the preselected state $\varrho$ can be seen as a quantifier of TC coherence in $\varrho$ relative to the eigenbasis $\{\ket{k}\}$ of $K$, or a quantifier of asymmetry relative to translation group generated by $K$, satisfying the above listed requirements expected for such a quantifier. Let us note that the concept of TC coherence or coherence as asymmetry was first discussed in Refs. \cite{Marvian coherence as asymmetry 0,Marvian coherence as asymmetry,Marvian - Spekkens speakable and unspeakable coherence}. It is argued in those papers within the quantum resource theory that any quantifier of TC coherence must satisfy the properties referenced in Propositions 1, 4, 6; hence, the free operations in the resource theory are given by the set of translationally-covariant quantum operations. As mentioned in Section \ref{Introduction}, TC coherence is also called unspeakable in Ref. \cite{Marvian - Spekkens speakable and unspeakable coherence} which means that its encoding depends on the specific elements of the incoherent basis appearing in the superposition. On the other hand, there is a different but closely related notion of speakable coherence \cite{Baumgratz quantum coherence measure,Streltsov review,Aberg quantifying of superposition,Levi quantum coherence measure} wherein the encoding is independent of the elements of the incoherent basis appearing in the superposition. From the viewpoint of resource theory, while both the speakable and unspeakable coherence have the same set of free states given by incoherent states having the form of (\ref{translationally symmetric state}), their set of free operations are different. Unlike the set of free operations in the unspeakable coherence or the TC coherence that is naturally given by the translationally-covariant quantum operations, there are a number of different proposals for the set of free operations for speakable coherence. See Ref. \cite{Marvian - Spekkens speakable and unspeakable coherence} for a detail comparison of the two concepts.    

We show that, unlike the quantifier of speakable coherence, the TC w-coherence defined in Eq. (\ref{TC w-coherence}) indeed depends on which elements in the eigenbasis $\{\ket{k}\}$ of $K$ appearing in the superposition. That is, it is not in general invariant under index permutation of the basis. To see this, consider the following transformation of basis: $\ket{k}\mapsto e^{i\theta_k}\ket{\mu(k)}$, where $\mu(k)$ is an index permutation and $\theta_k\in\mathbb{R}$. Such a basis transformation can thus be expressed by a unitary transformation as $\varrho\mapsto V_{\rm p}\varrho V_{\rm p}^{\dagger}$ where $V_{\rm p}=\sum_ke^{i\theta_k}\ket{\mu(k)}\bra{k}$. Inserting into the definition of TC w-coherence of Eq. (\ref{TC w-coherence}), we have $C_{\rm w}(V_{\rm p}\varrho V_{\rm p}^{\dagger};K)=\sup_{\{\ket{x'}\}}\sum_{x'}\big|{\rm Im}\braket{x'|V_{\rm p}^{\dagger}K V_{\rm p}\varrho|x'}\big|=C_{\rm w}(\varrho;V_{\rm p}^{\dagger}K V_{\rm p})$, where, we have again defined a new basis $\{\ket{x'}\}=\{V_p^{\dagger}\ket{x}\}$ and used $\sup_{\{\ket{x}\}}(\cdot)=\sup_{\{\ket{x'}\}}(\cdot)$. Note that $V_{\rm p}^{\dagger}K V_{\rm p}=\sum_k k\ket{\mu(k)}\bra{\mu(k)}=\sum_k\mu^{-1}(k)\ket{k}\bra{k}$, where we have used the spectral decomposition $K=\sum_kk\ket{k}\bra{k}$ and $\mu^{-1}(k)$ is the functional inverse of the permutation $\mu(k)$. Hence, while $K$ and $V_{\rm p}^{\dagger}K V_{\rm p}$ possess the same sets of eigenvalues and eigenvectors, the ordering of the eigenvectors relative to that of the eigenvalues are different due to the permutation. In general, we therefore have $V_{\rm p}^{\dagger}K V_{\rm p}\neq K$ so that 
\begin{eqnarray}
C_{\rm w}(V_{\rm p}\varrho V_{\rm p}^{\dagger};K)=C_{\rm w}(\varrho;V_{\rm p}^{\dagger}K V_{\rm p})\neq C_{\rm w}(\varrho;K). 
\label{TC w-coherence is not invariant under permutation}
\end{eqnarray}

Further remarks are in order. First, suppose that the generator of the translation group $K$ is degenerate. Hence, the eigenbasis of $K$ decomposes the Hilbert space into the direct sum of the subspaces $\mathcal{H}=\oplus_k\mathcal{H}_k$, with the associated projectors $\{\Pi_k\}$, i.e., $\mathcal{H}_k=\Pi_k\mathcal{H}$, so that not all of $\mathcal{H}_k$ is one-dimensional. In this case, the TC w-coherence defined in Eq. (\ref{TC w-coherence}) can be interpreted to quantify the TC quantum coherence relative to such decomposition of the Hilbert space. It is also clear that, as per definition, TC w-coherence cannot be applied to access coherence relative to nonorthogonal basis. Such a concept of coherence has found relevance in different physical situations \cite{Theurer coherence relative to non-orthogonal basis,Rastegin coherence relative to Luders and POVM measurement,Bischof coherence relative to POVM measurement,Das coherence relative to POVM measurement}. Finally, we note that when defining the TC w-coherence in Eqs. (\ref{TC w-coherence}) and (\ref{TC w-coherence for a composite}), we have deliberately chosen the weak value associated with projective measurement, i.e., those represented by projection-valued measure $\{\Pi_x\}$. This choice has allowed us to prove the nice properties discussed above. It is then natural to ask if one can extend the TC w-coherence in Eq. (\ref{TC w-coherence}) by using generalized weak value based on POVM \cite{Haapasalo generalized weak value,Busch book on quantum measurement}. This generalization will give a different quantity. It is however unclear if such generalization retains the plausible properties listed above.   

\subsection{Experimental schemes to determine TC w-coherence and its interpretation\label{Observation of TC w-coherence via the observation of weak value}}

To estimate the TC w-coherence of an unknown quantum state relative to the eigenbasis $\{\ket{k}\}$ of $K$, one can simply first make a quantum state tomography to get the density matrix $\varrho$, and compute $C_{\rm w}(\varrho;K)$ using the formula (\ref{TC w-coherence}). Such an approach however does not tell us the operational meaning of the TC w-coherence. Can we experimentally estimate the TC w-coherence of an unknown state without recoursing to the quantum state tomography? Or, can we translate the mathematical definition of TC w-coherence in Eq. (\ref{TC w-coherence}) directly into a set of laboratory operations and some classical processing? Fortunately, this can be done since the weak value itself, even though it is in general complex-valued, can be estimated in experiment, either using weak measurement with postselection as the weak value was originally conceived \cite{Aharonov weak value,Wiseman weak value,Dressel weak value review,Haapasalo generalized weak value,Lundeen complex weak value,Jozsa complex weak value}, or using a number of other methods \cite{Johansen quantum state from successive projective measurement,Johansen weak value from a sequence of strong measurement,Vallone strong measurement to reconstruct quantum wave function,Cohen estimating of weak value with strong measurements,Lostaglio KD quasiprobability and quantum fluctuation,Wagner measuring weak values and KD quasiprobability}.  

The general scheme for the direct estimation of the TC w-coherence in an unknown quantum state thus proceeds as follows. First, one estimates the imaginary part of the weak value $K^{\rm w}(\Pi_x|\varrho)$ via one of the schemes proposed in the literatures, with the input: the state $\varrho$ under scrutiny, the parameters that define the generator $K$ of the translation group, and the element of projective basis $\{\ket{x(\lambda_1,\dots,\lambda_M)}\}$, where $\{\lambda_i\}_{i=1}^M$ are the set of parameters that define the projective basis whose variation leads to scanning over all the bases of the relevant Hilbert space. See the next section for a concrete example of parameterization of such basis for a single qubit. This means that we need to have a parameterized unitary $U_{\{\lambda_1,\dots,\lambda_M\}}$ that can transform the standard basis to $\{\ket{x(\lambda_1,\dots,\lambda_M)}\}$. Next, one computes the average of the absolute imaginary part of the weak value over ${\rm Pr}(x(\lambda_1,\dots,\lambda_N)|\varrho)$ via a classical processing. Then, one varies the parameters $\{\lambda_i\}_{i=1}^N$ of the unitary $U_{\{\lambda_1,\dots,\lambda_M\}}$ that prepares the projective basis $\{\ket{x(\lambda_1,\dots,\lambda_N)}\}$, and repeat the procedure until one gets the converging supremum value. Hence, in this way, we have a hybrid quantum-classical procedure to estimate the TC w-coherence in the fashion of variational quantum circuit  \cite{Cerezo VQA review}, directly, without recoursing to the quantum state tomography. There are thus at least two problems in this protocol which are left for future investigation. One is the efficiency of the determination of the weak value and the other is the efficiency of the optimization scheme which is common to all variational quantum circuits.  

From the operational scheme to estimate the TC w-coherence, one may also suggest a physical interpretation of the TC w-coherence. For example, in the scheme based on weak measurement with postselection to obtain the weak value, the imaginary part of the weak value $K^{\rm w}(\Pi_x|\varrho)$ characterizes a disturbance of the state $\varrho$ induced by the unitary translation generated by $K$ \cite{Dressel imaginary weak value and disturbance}. In this sense, the TC w-coherence can thus be interpreted as the disturbance induced by the unitary translation maximized over all possible (postselection) bases $\{\ket{x}\}$ of the Hilbert space. This interpretation goes along with the intuition that the larger is the TC w-coherence in $\varrho$ relative to the translation group generated by $K$, the more sensitive is the state $\varrho$ under the unitary translation generated by $K$. This is the reason why larger TC coherence is desirable for quantum parameter estimation. On the other hand, in the scheme of obtaining weak value via a sequence of strong measurements proposed in Refs. \cite{Johansen quantum state from successive projective measurement,Johansen weak value from a sequence of strong measurement}, the imaginary part of the weak value arises as the disturbance of measurement of the incoherent basis $\{\ket{k}\}$ of K. In this context, the TC w-coherence thus measures the maximum of the total sum of such disturbance over all possible bases $\{\ket{x}\}$ of the Hilbert space. 

\subsection{TC w-coherence of a single qubit\label{Calculation for a single qubit}}

Let us give a concrete calculation of the TC w-coherence in a single qubit, i.e., two level system, with an arbitrary mixed state, relative to a translation group generated by an arbitrary Hermitian operator. Consider first the TC w-coherence relative to the eigenbasis of a Hermitian operator $K_z$ that is given by the eigenvectors of the Pauli $\sigma_z$ matrix: $\{\ket{0},\ket{1}\}$, with the corresponding eigenvalues $\{k_0,k_1\}$, so that $K_z=k_0\ket{0}\bra{0}+k_1\ket{1}\bra{1}$. It is convenient to parameterize the basis $\{\ket{x(\alpha,\beta)}\}=\{\ket{x_+(\alpha,\beta)},\ket{x_-(\alpha,\beta)}\}$ of the two-dimensional Hilbert space in the Bloch sphere as:
\begin{eqnarray}
\ket{x_+(\alpha,\beta)}&:=&\cos\big(\frac{\alpha}{2}\big)\ket{0}+\sin\big(\frac{\alpha}{2}\big)e^{i\beta}\ket{1};\nonumber\\
\ket{x_-(\alpha,\beta)}&:=&\sin\big(\frac{\alpha}{2}\big)\ket{0}-\cos\big(\frac{\alpha}{2}\big)e^{i\beta}\ket{1}, 
\label{complete set of basis in the x-y plane}
\end{eqnarray}
$0\le\alpha\le\pi$, $0\le\beta < 2\pi$. Hence, one can scan over all the possible bases of the two-dimensional Hilbert space by varying the angular parameters $\alpha$ and $\beta$ over the whole ranges of values. Using this expression for the defining basis in Eq. (\ref{TC w-coherence}), and writing the general qubit state as $\varrho=\sum_{i,j}\varrho_{ij}\ket{i}\bra{j}$, $\varrho_{ij}=\braket{i|\varrho|j}$,  $i,j={0,1}$, we directly get
\begin{eqnarray}
C_{\rm w}(\varrho;K_z)&=&\sup_{\{\ket{x(\alpha,\beta)}\}}\sum_x|{\rm Im}\braket{x(\alpha,\beta)|K_z\varrho|x(\alpha,\beta)}|\nonumber\\
&=&\max_{\alpha,\beta}|k_0-k_1||\varrho_{01}||\sin\alpha||\sin(\beta+\phi_{01})|\nonumber\\
&=&|k_0-k_1||\varrho_{01}|,
\label{TC w-coherence for a qubit}
\end{eqnarray}
where $\phi_{01}=\arg(\varrho_{01})$, and the maximum is obtained for the basis of Eq. (\ref{complete set of basis in the x-y plane}) with $\alpha=\pi/2$ and $\beta=\pi/2-\phi_{01}$. Notice that when $K_z$ is degenerate, i.e., $k_0=k_1$, we have $C_{\rm w}(\varrho;K_z)=0$ for all $\varrho$, as desired. Moreover, one also finds that the TC w-coherence in the state relative to the eigenbasis $\{\ket{0},\ket{1}\}$ of $K_z$, captures the magnitude of the off-diagonal terms of the density matrix $\varrho$ in the basis $\{\ket{0},\ket{1}\}$, as intuitively expected. Hence, for a single qubit, the TC w-coherence relative to the translation group generated by $K_z$ is formally proportional to the $l_1$-norm coherence with respect to the incoherent basis $\{\ket{0},\ket{1}\}$ which is just given by $C_{l_1}(\varrho;\{\ket{0},\ket{1}\})=2|\varrho_{01}|$ \cite{Baumgratz quantum coherence measure}. For the specific case when $k_0=-k_1=1$ so that $K_z=\sigma_z$, they are exactly equal: $C_{\rm w}(\varrho;\sigma_z)=C_{l_1}(\varrho;\{\ket{0},\ket{1}\})=2|\varrho_{01}|$. 

The above result can be generalized to TC w-coherence relative to the eigenbasis $\{\ket{k_+},\ket{k_-}\}$ of any nondegenerate Hermitian operator $K=k_+\ket{k+}\bra{k+}+k_-\ket{k_+}\bra{k_-}$ on two-dimensional Hilbert space with the real eigenvalues $\{k_+,k_-\}$, $k_+\neq k_-$. To see this, we recall that according to Proposition 3, TC w-coherence is unitarily covariant 
\begin{eqnarray}
C_{\rm w}(\varrho;K)=C_{\rm w}(\varrho';K'), 
\label{unitarily covariant for a qubit}
\end{eqnarray}
where $\varrho'=V\varrho V^{\dagger}$ and $K'=VKV^{\dagger}$, and $V$ is any unitary transformation. Let us choose the following unitary transformation: $V=\ket{0}\bra{k_+}+\ket{1}\bra{k_-}$ so that we have $K'=VKV^{\dagger}=k_+\ket{0}\bra{0}+k_-\ket{1}\bra{1}$, and $\varrho'=V\varrho V^{\dagger}=\varrho_{++}\ket{0}\bra{0}+\varrho_{+-}\ket{0}\bra{1}+\varrho_{-+}\ket{1}\bra{0}+\varrho_{--}\ket{1}\bra{1}$, where $\varrho_{\pm\pm}=\braket{k_{\pm}|\varrho|k_{\pm}}$. Inserting these into Eq. (\ref{unitarily covariant for a qubit}), and noting Eq. (\ref{TC w-coherence for a qubit}), we thus obtain 
\begin{eqnarray}
C_{\rm w}(\varrho;K)=C_{\rm w}(\varrho';K')=|k_+-k_-||\varrho_{+-}|. 
\label{TC w-coherence with respect to generic observable for a qubit}
\end{eqnarray} 
Hence, for $k_+=-k_-=1$, we again have $C_{\rm w}(\varrho;K)=C_{l_1}(\varrho;\{\ket{k_+},\ket{k_-}\})=2|\varrho_{+-}|$. It is also  clear from the above result that for two-dimensional system, the TC w-coherence is invariant under index permutation of the reference basis. We stress that the above equality for the TC w-coherence and $l_1$-norm coherence cannot in general be maintained for qudit with $d\ge 3$. In particular, while $l_1$-norm, being a measure for speakable coherence, is invariant under permutation of the elements of the incoherent basis, as shown in Eq. (\ref{TC w-coherence is not invariant under permutation}), TC w-coherence is in general not invariant under such transformation.  

A comment on the calculation of the TC w-coherence defined in Eqs. (\ref{TC w-coherence}) or (\ref{TC w-coherence for a composite}) in a quantum state on Hilbert space with dimension larger than two is in order. We first note that to compute the TC w-coherence, we need a parameterization of the defining basis using in general multivariable parameters, $\{\ket{x(\lambda_1,\dots,\lambda_M)}\}$. The parameters $\{\lambda_1,\dots,\lambda_M\}$ are then varied to scan over all the possible bases of the respected Hilbert space to obtain the supremum in Eqs. (\ref{TC w-coherence}) or (\ref{TC w-coherence for a composite}). For example, for a composite system of $N$ qubits, to compute the TC w-coherence relative to the translation group generated by a local Hermitian operator, one needs to vary $2N$ real variables parameterizing the defining product basis. Hence, in general, the computation of TC w-coherence is analytically intractable involving optimization of multivariable nonlinear function. On the other hand, there are quantifiers of coherence as asymmetry in the literature which can be computed directly given the quantum state $\varrho$ and via diagonalization, such as that based on Wigner-Yanase skew information \cite{Wigner-Yanase skew information} or the 1-norm of the commutation relation between the state $\varrho$ and the generator $K$ of the translational group \cite{Marvian - Spekkens speakable and unspeakable coherence,Girolami quantum coherence measure}. Note however that these quantifiers do not translate directly to laboratory operations combined with classical data processing. Hence, one needs to first make a full quantum state tomography, rendering the operational meaning of these quantifiers not entirely clear. By contrast, as discussed in Section \ref{Observation of TC w-coherence via the observation of weak value}, the definition of TC w-coherence in Eqs. (\ref{TC w-coherence}) or (\ref{TC w-coherence for a composite}) translates directly to laboratory operations in terms of the estimation of weak values albeit combined with a classical optimization procedure. 

\section{TC w-coherence and quantum statistics}

\subsection{TC w-coherence and quantum standard deviation\label{TC w-coherence and quantum standard deviation}}

Notice that  the TC w-coherence $C_{\rm w}(\varrho;K)$ defined in Eq. (\ref{TC w-coherence}) expresses the extent of the noncommutativity between the generator $K$ of the translation group and the quantum state $\varrho$ under scrutiny, optimized over all the bases of the associated Hilbert space. It should therefore be related with the quantum uncertainty in the measurement of Hermitian operator $K$ over $\varrho$, which also partially arises from the noncommutativity between $K$ and $\varrho$. Hence, it is instructive to compare the TC w-coherence with the quantum variance of $K$ in $\varrho$, i.e., the standard quantifier of the total uncertainty arising in the measurement of the observable $K$ over $\varrho$ which also includes the classical uncertainty due to the statistical mixing of quantum states. We have the following theorem.\\
{\bf Theorem 1}. The TC w-coherence in the quantum state $\varrho$ relative to the eigenbasis $\{\ket{k}\}$ of the generator $K$ of a translation group is always less than or equal to the quantum standard deviation of the observable $K$ in $\varrho$, i.e.:
\begin{eqnarray}
C_{\rm w}[\varrho ;K]\le\Delta_K(\varrho), 
\label{weak coherence versus quantum uncertainty}
\end{eqnarray}  
where $\Delta^2_K(\varrho ):={\rm Tr}\{K^2\varrho \}-({\rm Tr}\{K\varrho \})^2$ is the quantum variance of $K$ in the state $\varrho$. \\
{\bf Proof}. First, we have, from Eq. (\ref{TC w-coherence}) and using the Jensen inequality, 
\begin{eqnarray}
C_{\rm w}(\varrho;K)&=&\sup_{\{\ket{x}\}}\sum_x\big|{\rm Im}(K^{\rm w}(\Pi_x|\varrho))\big|{\rm Tr}(\Pi_x\varrho)\nonumber\\
&\le&\sup_{\{\ket{x}\}}\big(\sum_x({\rm Im}K^{\rm w}(\Pi_x|\varrho))^2{\rm Tr}(\Pi_x\varrho)\big)^{1/2}. 
\label{TC w-coherence and MSE of estimation of generator}
\end{eqnarray}
Next, noting that $({\rm Im}K^{\rm w}(\Pi_x|\varrho))^2=|K^{\rm w}(\Pi_x|\varrho)|^2-({\rm Re}K^{\rm w}(\Pi_x|\varrho))^2$, and inserting into Eq. (\ref{TC w-coherence and MSE of estimation of generator}), we obtain, upon using the definition of the weak value of Eq. (\ref{complex weak value}), 
\begin{eqnarray}
&&C_{\rm w}(\varrho;K)\nonumber\\
\label{from weak measurement to quantum uncertainty step 0}
&\le&\Big[\sum_{x_*}\Big(\Big|\frac{{\rm Tr}(\Pi_{x_*}K\varrho)}{{\rm Tr}(\Pi_{x_*}\varrho)}\Big|^2-{\rm Re}\Big(\frac{{\rm Tr}(\Pi_{x_*}K\varrho)}{{\rm Tr}(\Pi_{x_*}\varrho)}\Big)^2\Big){\rm Tr}(\Pi_{x_*}\varrho)\Big]^{1/2}\\
\label{from weak measurement to quantum uncertainty step 1}
&\le&\Big[\sum_{x_*}\frac{|{\rm Tr}(\Pi_{x_*}K\varrho)|^2}{{\rm Tr}(\Pi_{x_*}\varrho)}-\big(\sum_{x_*}{\rm Re}\big({\rm Tr}(\Pi_{x_*}K\varrho)\big)\big)^2\Big]^{1/2},
\end{eqnarray}
where Eq. (\ref{from weak measurement to quantum uncertainty step 0}) holds since $\{\ket{x_*}\}$ is the basis which achieves the supremum, and Eq. (\ref{from weak measurement to quantum uncertainty step 1}) holds due to the Jensen inequality, i.e., $(\sum_{x_*}{\rm Re}\{{\rm Tr}\{\Pi_{x_*}K\varrho\}\})^2=(\sum_{x_*}\frac{{\rm Re}\{{\rm Tr}\{\Pi_{x_*}K\varrho\}\}}{{\rm Tr}\{\Pi_{x^*}\varrho\}}{\rm Tr}\{\Pi_{x^*}\varrho\})^2\le\sum_{x_*}(\frac{{\rm Re}\{{\rm Tr}\{\Pi_{x_*}K\varrho\}\}}{{\rm Tr}\{\Pi_{x^*}\varrho\}})^2{\rm Tr}\{\Pi_{x^*}\varrho\}$. Finally, applying the Cauchy-Schwartz inequality, i.e., $|{\rm Tr}(A^{\dagger}B)|^2\le{\rm Tr}(A^{\dagger}A){\rm Tr}(B^{\dagger}B)$, to the numerator in the first term on the right-hand side of Eq. (\ref{from weak measurement to quantum uncertainty step 1}): $|{\rm Tr}(\Pi_{x_*}K\varrho)|^2=|{\rm Tr}((\Pi_{x_*}^{1/2}K\varrho ^{1/2})(\varrho ^{1/2}\Pi_{x_*}^{1/2}))|^2\le{\rm Tr}(\Pi_{x_*}K\varrho K){\rm Tr}(\Pi_{x_*}\varrho)$, and using the completeness relation $\sum_{x_*}\Pi_{x_*}=\mathbb{I}$, we obtain Eq. (\ref{weak coherence versus quantum uncertainty}). \qed

Let us evaluate the right-hand side of Eq. (\ref{weak coherence versus quantum uncertainty}) for a single qubit. First, we consider the case when the generator of the translation takes the following form: $K_z=k_0\ket{0}\bra{0}+k_1\ket{1}\bra{1}$, $k_0\neq k_1$. For later purpose, it is convenient to express the general state of the qubit as $\varrho=\frac{1}{2}(\mathbb{I}+r_x\sigma_x+r_y\sigma_y+r_z\sigma_z)$, where $\sigma_x,\sigma_y,\sigma_z$ are the Pauli operators, and $r^2=r_x^2+r_y^2+r_z^2\le 1$. One then gets 
\begin{eqnarray}
\Delta_{K_z}^2(\varrho)=\frac{1}{4}(k_0-k_1)^2(1-r_z^2). 
\label{variance of an observable in a qubit}
\end{eqnarray}
On the the other hand, for a single qubit, noting Eq. (\ref{TC w-coherence for a qubit}) and the fact that $\varrho_{01}=\frac{1}{2}(r_x-ir_y)$, the square of the TC w-coherence relative to the eigenbasis $\{\ket{0},\ket{1}\}$ of $K_z$ can be expressed as:
\begin{eqnarray}
C_{\rm w}(\varrho;K_z)^2&=&\frac{1}{4}(k_0-k_1)^2(r_x^2+r_y^2)=\frac{1}{4}(k_0-k_1)^2(r^2-r_z^2)\nonumber\\
&\le&\frac{1}{4}(k_0-k_1)^2(1-r_z^2)=\Delta_{K_z}^2(\varrho),
\label{TC w-coherence vs quantum variance for a single qubit}
\end{eqnarray} 
where the last equality is just Eq. (\ref{variance of an observable in a qubit}). Notice that it satisfies the inequality in Eq. (\ref{weak coherence versus quantum uncertainty}), and for pure states where $r=1$, the inequality in Eq. (\ref{TC w-coherence vs quantum variance for a single qubit}) becomes equality.  

The above result can be generalized to an arbitrary Hermitian operator $K=k_+\ket{k_+}\bra{k_+}+k_-\ket{k_-}\bra{k_-}$, $k_+\neq k_-$. We note first that the quantum variance of an arbitrary observable $K$ in any state $\varrho$ is unitarily covariant, i.e., $\Delta_K^2(\varrho)=\Delta^2_{VKV^{\dagger}}(V\varrho V^{\dagger})$, for any unitary transformation $V$. We again choose $V=\ket{0}\bra{k_+}+\ket{1}\bra{k_-}$ so that $K'=VKV^{\dagger}=k_+\ket{0}\bra{0}+k_-\ket{1}\bra{1}$. Writing $V\varrho V^{\dagger}=\frac{1}{2}(\mathbb{I}+r'_x\sigma^x+r'_y\sigma^y+r'_z\sigma^z)=\varrho'$, where ${r'_x}^2+{r'_y}^2+{r'_z}^2=r'^2=r^2$ (due to conservation of the purity of state under unitary transformation), we thus obtain, noting Eq. (\ref{variance of an observable in a qubit}): $\Delta_K^2(\varrho)=\Delta_{K'}^2(\varrho')=\frac{1}{4}(k_+-k_-)^2(1-{r'_z}^2)$. On the other hand, in this case, the TC w-coherence in $\varrho$ relative to the eigenbasis $\{\ket{k_+},\ket{k_-}\}$ of $K$ is given by Eq. (\ref{TC w-coherence with respect to generic observable for a qubit}). We thus finally have, in accord with Eq. (\ref{weak coherence versus quantum uncertainty}),
\begin{eqnarray}
C_{\rm w}(\varrho;K)^2&=&(k_+-k_-)^2|\varrho_{+-}|^2\nonumber\\
&=&\frac{1}{4}(k_+-k_-)^2(r^2-{r'_z}^2)\nonumber\\
&\le&\frac{1}{4}(k_+-k_-)^2(1-{r'_z}^2)=\Delta_K^2(\varrho),
\label{TC w-coherence vs quantum variance for a single qubit: generic}
\end{eqnarray}
where the equality is again reached for pure states with $r=1$. We thus obtain the following result. \\
{\bf Theorem 2}. For an arbitrary pure state of a single qubit, the square of the TC w-coherence in $\varrho$ relative to the eigenbasis $\{\ket{k}\}$ of $K$ is exactly equal to the quantum variance of $K$ in $\varrho$ saturating Eq. (\ref{weak coherence versus quantum uncertainty}).  

The above observation suggests that the TC w-coherence $C_{\rm w}(\varrho;K)$ defined in Eq. (\ref{TC w-coherence}) may be seen statistically as capturing a genuine quantum uncertainty out of the total uncertainty, arising in the measurement of the observable $K$ over the state $\varrho$ due to their noncommutativity. Note that any quantity capturing the genuine quantum uncertainty of $K$ in $\varrho$ must satisfy the following plausible properties: (i) vanishing if and only if $K$ and $\varrho$ commute, (ii) convex (i.e., nonincreasing under classical mixing in $\varrho$), (iii) upper bounded by the standard quantum deviation of $K$ in $\varrho$ for general state, and (iv) equal to the quantum standard deviation of $K$ in $\varrho$ for all pure states $\varrho=\ket{\psi}\bra{\psi}$. We have shown above that the TC w-coherence $C_{\rm w}(\varrho;K)$ satisfies the requirements (i)-(iii), and it satisfies (iv) for a single qubit. It is an open problem whether TC w-coherence satisfies (iv) for arbitrary finite dimension of Hilbert space. Such a genuine quantum uncertainty of $K$ in $\varrho$ is also called ``coherent spread'' of the state $\varrho$ over the eigenbasis of the $K$ \cite{Marvian - Spekkens speakable and unspeakable coherence}.   

\subsection{TC w-coherence and quantum Fisher information\label{TC w-coherence and quantum Fisher information}}

Consider again the protocol of quantum parameter estimation wherein an unknown parameter $\theta$ is imprinted to the state $\varrho_{\theta}$ of the probe via a unitary translation $U_{K,\theta}$ of Eq. (\ref{unitary translation}). Now, one wishes to estimate $\theta$ from the statistics of the outcomes of the most general measurement allowed by quantum mechanics described by a set of POVM $\{M_x\}$, $0\le M_x\le \mathbb{I}$, $\sum_xM_x=\mathbb{I}$, with the outcomes $\{x\}$ and probability ${\rm Pr}(x|\varrho_{\theta})={\rm Tr}\{M_x\varrho_{\theta}\}$. When the estimation is locally unbiased, its sensitivity is characterized by the quantum Cram\'er-Rao inequality as \cite{Helstrom estimation-based UR,Holevo book on quantum statistics}
\begin{eqnarray}
\Delta^2_{\hat{\theta}}(\varrho_{\theta})\ge\frac{1}{\nu \mathcal{J}_{\theta}(\varrho_{\theta})}.
\label{quantum Cramer-Rao inequality}
\end{eqnarray}
Here, $\nu$ is the number of probing, and $\mathcal{J}_{\theta}(\varrho_{\theta})$ is the quantum Fisher information defined as follows.\\
{\bf Definition 4}. The quantum Fisher information about a scalar parameter $\theta$ contained in a quantum state $\varrho_{\theta}$ is given by \cite{Braunstein estimation-based UR 1,Braunstein estimation-based UR 2,Paris quantum estimation review}: 
\begin{eqnarray}
\mathcal{J}_{\theta}(\varrho_{\theta})&:=&\sup_{\{M_x\}}J_{\theta}(\{{\rm Pr}(x|\varrho_{\theta})\}), 
\label{quantum Fisher information}
\end{eqnarray}
where $J_{\theta}(\{{\rm Pr}(x|\varrho_{\theta})\})$ is the classical Fisher information about $\theta$ contained in ${\rm Pr}(x|\varrho_{\theta})$ defined in Eq. (\ref{classical Fisher information}), and the supremum is taken over all POVM $\{M_x\}$. \\
Noting this, we then have the following theorem. \\
{\bf Theorem 3}. The square of the TC w-coherence in $\varrho_{\theta}$ relative to the eigenbasis $\{\ket{k}\}$ of the generator $K$ of translation group, is upper bounded by the quantum Fisher information about $\theta$ in $\varrho_{\theta}$ obtained via a unitary imprinting generated by $K$ as 
\begin{eqnarray}
C_{\rm w}(\varrho_{\theta};K)^2\le\mathcal{J}_{\theta}(\varrho_{\theta})/4.
\label{TC w-coherence is upper bounded by quantum Fisher information}
\end{eqnarray}
{\bf Proof}. First, from the definition of the TC w-coherence of Eq. (\ref{TC w-coherence}), and using Eq. (\ref{imaginary part of weak value and score function}), we can express the TC w-coherence as 
\begin{eqnarray}
C_{\rm w}(\varrho_{\theta};K)=\frac{1}{2}\sup_{\{\Pi_x\}}\sum_x\Big|\frac{\partial_{\theta}{\rm Pr}(x|\varrho_{\theta})}{{\rm Pr}(x|\varrho_{\theta})}\Big|{\rm Pr}(x|\varrho_{\theta}). 
\label{TC w-coherence as average score function}
\end{eqnarray}
The inequality of Eq. (\ref{TC w-coherence is upper bounded by quantum Fisher information}) can then be derived as 
\begin{eqnarray}
\mathcal{J}_{\theta}(\varrho_{\theta})&:=&\sup_{\{M_x\}}\sum_x\Big|\frac{\partial_{\theta}{\rm Pr}(x|\varrho_{\theta})}{{\rm Pr}(x|\varrho_{\theta})}\Big|^2{\rm Pr}(x|\varrho_{\theta})\\
\label{TC w-coherence as the lower bound to the quantum Fisher information step 1}
&\ge&\sup_{\{\Pi_x\}}\sum_x\Big|\frac{\partial_{\theta}{\rm Pr}(x|\varrho_{\theta})}{{\rm Pr}(x|\varrho_{\theta})}\Big|^2{\rm Pr}(x|\varrho_{\theta})\\
\label{TC w-coherence as the lower bound to the quantum Fisher information step 2}
&\ge&\sup_{\{\Pi_x\}}\Big(\sum_x\Big|\frac{\partial_{\theta}{\rm Pr}(x|\varrho_{\theta})}{{\rm Pr}(x|\varrho_{\theta})}\Big|{\rm Pr}(x|\varrho_{\theta})\Big)^2\\ 
\label{TC w-coherence as the lower bound to the quantum Fisher information step 3}
&=&4C_{\rm w}(\varrho_{\theta};K)^2
\end{eqnarray} 
Here, the inequality in Eq. (\ref{TC w-coherence as the lower bound to the quantum Fisher information step 1}) holds since the set of POVM $\{M_x\}$ is larger than the set of projection-valued measure $\{\Pi_x\}$, the inequality in Eq. (\ref{TC w-coherence as the lower bound to the quantum Fisher information step 2}) is due the Jensen inequality, and the equality in Eq. (\ref{TC w-coherence as the lower bound to the quantum Fisher information step 3}) is just the expression of the TC w-coherence in Eq. (\ref{TC w-coherence as average score function}). \qed 

As an implication of the Theorem 2 we obtain the following corollary which is just a special case of Theorem 3.\\
{\bf Corollary 1}. For a pure state single qubit, the inequality of Eq. (\ref{TC w-coherence is upper bounded by quantum Fisher information}) is saturated.  \\
{\bf Proof}. First, it is known that for pure state on arbitrary finite dimension of the Hilbert space the quantum Fisher information $\mathcal{J}_{\theta}(\varrho_{\theta})$ is proportional to the quantum variance of the generator $K$ of the unitary shift along $\theta$ as $\mathcal{J}_{\theta}(\varrho_{\theta})=4\Delta^2_K(\varrho_{\theta})$ \cite{Braunstein estimation-based UR 1,Braunstein estimation-based UR 2,Paris quantum estimation review}. On the other hand, from Theorem 2, for a single qubit with pure state, we have $C_{\rm w}(\varrho_{\theta};K)^2=\Delta^2_K(\varrho_{\theta})$. Combining these two equalities, we indeed obtain $C_{\rm w}(\varrho_{\theta};K)^2=\mathcal{J}_{\theta}(\varrho_{\theta})/4$.  \qed

As an application of the above result, let us consider the optimal estimation scheme wherein the quantum Cram\'er-Rao bound of Eq. (\ref{quantum Cramer-Rao inequality}) is saturated. Then, the optimal mean-squared error of the estimate is bounded from above by the inverse of the squared TC w-coherence in the initial state $\varrho$ relative to the translation group generated by $K$ as   
\begin{eqnarray}
\label{coherence as the resource for the quantum phase estimate step 1}
\Delta^2_{\theta_{\rm opt}}(\varrho_{\theta})&:=&\frac{1}{\nu\mathcal{J}_{\theta}(\varrho_{\theta})}\le\frac{1}{4\nu C_{\rm w}(\varrho_{\theta};K)^2}\\
\label{coherence as the resource for the quantum phase estimate step 2}
&=&\frac{1}{4\nu C_{\rm w}(\varrho;K)^2}.  
\end{eqnarray}
Here, the inequality in Eq. (\ref{coherence as the resource for the quantum phase estimate step 1}) holds due to Eq. (\ref{TC w-coherence is upper bounded by quantum Fisher information}), and to get the equality in Eq. (\ref{coherence as the resource for the quantum phase estimate step 2}) we have used the Proposition 4 in Section \ref{Quantifier of TC coherence from the imaginary part of weak value}, i.e., $C_{\rm w}(\varrho_{\theta};K)=C_{\rm w}[U_{K,\theta}\varrho U_{K,\theta}^{\dagger};K]=C_{\rm w}(\varrho;K)$. In this sense, the TC w-coherence in the initial quantum state $\varrho$ of the probe, relative to the eigenbasis $\{\ket{k}\}$ of $K$ generating the unitary translation imprinting $\theta$ to the state of the probe, is a sufficient resource to guarantee an optimal achievable accuracy of the quantum parameter estimation. A similar result is obtained based on Wigner-Yanase skew information \cite{Wigner-Yanase skew information}  as a measure of discord-like general quantum correlation \cite{Girolami quantum correlation based on WY skew information}, by showing that the latter also gives a lower bound to the quantum Fisher information  \cite{Luo Wigner-Yanase skew information vs Fisher information}.  

\subsection{TC w-coherence and nonclassical-complex Kirkwood-Dirac quasiprobability \label{TC w-coherence, KD quasiprobability and KWR lower bound}}

Let us discuss the connection between the TC w-coherence and the KD (Kirkwood-Dirac) quasiprobability, whose nonclassical values (the meaning of which to be clarified below) also capture the noncommutativity between the corresponding quantum state and the defining bases. First, the KD quasiprobability is defined as follows \cite{Kirkwood quasiprobability,Dirac quasiprobability,Chaturvedi KD distribution}.\\
{\bf Definition 5}. Given a quantum state $\varrho$ on a Hilbert space, the associated KD quasiprobability  over a pair of orthonormal bases $\{\ket{k}\}$ and $\{\ket{x}\}$ of the Hilbert space, is defined as 
\begin{eqnarray}
{\rm Pr}_{\rm KD}(k,x|\varrho):={\rm Tr}(\Pi_x\Pi_k\varrho).
\label{KD quasiprobability}
\end{eqnarray} 

KD quasiprobability returns correct marginal probabilities, i.e., $\sum_x{\rm Pr}_{\rm KD}(k,x|\varrho)={\rm Tr}\{\Pi_k\varrho\}={\rm Pr}(k|\varrho)$ and $\sum_k{\rm Pr}_{\rm KD}(k,x|\varrho)={\rm Tr}\{\Pi_x\varrho\}={\rm Pr}(x|\varrho)$. It is one of the quantum analogs of phase space probability distribution in classical statistical mechanics. However, because of the quantum noncommutativity, the Kirkwood-Dirac quasiprobability may assume complex value and its real part may be negative. In this sense, the negativity or/and the nonreality of the Kirkwood-Dirac quasiprobability captures a form of nonclassicality and called KD nonclassicality. Indeed, the KD quasiprobability arises naturally in different forms of quantum fluctuations, and KD nonclassicality has been argued to signify genuine quantum behaviour of the underlying physical processes \cite{Lostaglio KD quasiprobability and quantum fluctuation}. Moreover, it is formally related to weak value: the weak value of a projector can be seen as the conditional KD quasiprobability. Hence, like weak value, it plays important roles in diverse fields of quantum science and information \cite{Lundeen direct measurement of wave function,Lundeen measurement of KD distribution,Lostaglio contextuality in quantum linear response,Allahverdyan TMH as quasiprobability distribution of work,Levy quasiprobability distribution for heat fluctuation in quantum regime,Halpern quasiprobability and information scrambling,Alonso KD quasiprobability witnesses quantum scrambling,Lostaglio KD quasiprobability and quantum fluctuation,Pusey strange weak value and contextuality,Lostaglio TMH quasiprobability fluctuation theorem contextuality,Kunjwal contextuality of non-real weak value,Levy quasiprobability distribution for heat fluctuation in quantum regime}. It is thus instructive to study the relation between the KD quasiprobability and the TC w-coherence defined in Eq. (\ref{TC w-coherence}) in terms of nonreal weak values.  

First, using the KD quasiprobability, the TC w-coherence in a quantum state $\varrho$ relative to the eigenbasis $\{\ket{k}\}$ of the generator $K$ of the translation group is upper bounded as follows 
\begin{eqnarray}
C_{\rm w}(\varrho;K)
&=&\sup_{\{\ket{x}\}}\sum_x\big|\sum_kk{\rm Im}{\rm Tr}(\Pi_x\Pi_k\varrho)\big|\nonumber\\
&\le&|k|_{\rm max}\sup_{\{\ket{x}\}}\sum_{k,x}\big|{\rm Im}{\rm Pr}_{\rm KD}(k,x|\varrho)\big|.
\label{TC w-coherence vs imaginary KD quasiprobability 1}
\end{eqnarray}
Here, we have used the spectral decomposition $K=\sum_kk\Pi_k$, with $|k|_{\rm max}=\max\{|k|\}$ is the maximum singular value, i.e., the spectral radius, of $K$. The above result suggests the following definition of normalized TC w-coherence. \\
{\bf Definition 6}. The normalized TC w-coherence in a quantum state $\varrho$ relative to the translation group generated by a Hermitian operator $K$ is defined as the associated TC w-coherence divided by the spectral radius of the generator:
\begin{eqnarray}
\tilde{C}_{\rm w}(\varrho;K):=C_{\rm w}(\varrho;K)/|k|_{\rm max}=C_{\rm w}(\varrho;\tilde{K}),
\label{normalized TC w-coherence}
\end{eqnarray}
where $\tilde{K}:=K/|k|_{\rm max}=\sum_k(k/|k|_{\rm max})\ket{k}\bra{k}$ is the generator of the translation group rescaled by its spectral radius. \\ 
We then obtain the following theorem. \\
{\bf Theorem 4}. The normalized TC w-coherence in $\varrho$ relative to the translational group generated $K$ is upper bounded by the total sum of the absolute imaginary part of the associated KD quasiprobability defined over the eigenbasis $\{\ket{k}\}$ of $K$, and a second basis $\{\ket{x}\}$, maximized over all the second bases of the Hilbert space:
\begin{eqnarray}
\tilde{C}_{\rm w}(\varrho;K)\le\sup_{\{\ket{x}\}}\sum_{k,x}\big|{\rm Im}{\rm Pr}_{\rm KD}(k,x|\varrho)\big|.
\label{normalized TC is upper bounded by the total sum of imaginary part of KD quasiprobability}
\end{eqnarray}
{\bf Proof}. Using the definition of the normalized TC w-coherence of Eq. (\ref{normalized TC w-coherence}) in Eq. (\ref{TC w-coherence vs imaginary KD quasiprobability 1}) we directly get Eq. (\ref{normalized TC is upper bounded by the total sum of imaginary part of KD quasiprobability}). \qed\\
Theorem 4 shows that the nonclassicality captured by the TC w-coherence is deeply related to the nonclassicality captured by the KD quasiprobability. 

Now, consider a set $\Lambda_{\{k\}}$ of all possible Hermitian operators $K$ having the same nontrivial eigenvalues spectrum $\{k\}$. Then, maximizing Eq. (\ref{normalized TC is upper bounded by the total sum of imaginary part of KD quasiprobability}) over all $K\in\Lambda_{\{k\}}$, we have the following corollary of Theorem 4. \\ 
{\bf Corollary 2}. The maximum normalized TC w-coherence in $\varrho$ relative to the translational groups generated by all possible $K\in\Lambda_{\{k\}}$ having the same nontrivial eigenvalues spectrum $\{k\}$, is upper bounded by the total sum of the absolute imaginary part of the associated KD quasiprobability maximized over all possible pair of the defining bases:
\begin{eqnarray}
\sup_{K\in\Lambda_{\{k\}}}\tilde{C}_{\rm w}(\varrho;K)\le \sup_{\{\ket{k};\ket{x}\}}\sum_{k,x}\big|{\rm Im}{\rm Pr}_{\rm KD}(k,x|\varrho)\big|.
\label{TC w-coherence vs imaginary KD quasiprobability 2}
\end{eqnarray}   

\subsection{Uncertainty relation for TC w-coherences\label{Uncertainty relation for TC w-coherences}}

We first derive a lower bound for the TC w-coherence as follows. Let us consider a set $\Lambda_{\{x\}}$ of all Hermitian operators $X$ on the Hilbert space of the system having a nontrivial spectrum of eigenvalues $\{x\}$. Then we have
\begin{eqnarray}
C_{\rm w}(\varrho;K)&=&\sup_{\{\ket{x}\}}\sum_{x}|{\rm Im}{\rm Tr}(\Pi_xK\varrho)|\nonumber\\
&=&\sup_{\{\ket{x}\}}\frac{1}{|x|_{\rm max}}\sum_{x}|x|_{\rm max}|{\rm Im}{\rm Tr}(\Pi_{x}K\varrho)|\nonumber\\
&\ge&\sup_{X\in\Lambda_{\{x\}}}\frac{1}{2|x|_{\rm max}}|{\rm Tr}([X,K]\varrho)|,
\label{coherence and incompatibility 0}
\end{eqnarray}
where we have used $X:=\sum_{x}x\Pi_x$, and $|x|_{\rm max}=\max\{|x|\}$ is the spectral radius of $X$. We thus obtain the following lemma by combining Eqs. (\ref{normalized TC w-coherence}) and (\ref{coherence and incompatibility 0}).\\
{\bf Lemma 1}. The normalized TC w-coherence in $\varrho$ relative to the translation group generated by $K$ is lower bounded by the maximum average noncommutativity between the generator of the translation $K$ and any other possible Hermitian operators $X\in\Lambda_{\{x\}}$ whose eigenbasis spans the Hilbert space as 
\begin{eqnarray}
\tilde{C}_{\rm w}(\varrho;K)\ge\sup_{X\in\Lambda_{\{x\}}}\frac{1}{2}|{\rm Tr}([\tilde{X},\tilde{K}]\varrho)|,
\label{coherence and incompatibility 1}
\end{eqnarray}
where $\tilde{X}:=X/|x|_{\rm max}$. 

Notice that the term on the right-hand side, i.e., $\frac{1}{2}|{\rm Tr}\{[\tilde{X},\tilde{K}]\varrho\}|$, is the lower bound of the KWR (Kennard-Weyl-Robertson) uncertainty relation for $K$ and $X$ over the quantum state $\varrho$, normalized by the spectral radiuses of $X$ and $K$. The above result can also be read as follows. Suppose we have a state $\varrho$ and a Hermitian operator $K$ and we wish to maximize the normalized KWR lower bound $\frac{1}{2}|{\rm Tr}\{[\tilde{X},\tilde{K}]\varrho\}|$, by varying the Hermitian operator $X$ over a certain set $\Lambda_{\{x\}}$ with a fixed spectrum $\{x\}$. Then, Eq. (\ref{coherence and incompatibility 1}) shows that the maximum value is upper bounded by the normalized TC w-coherence in $\varrho$ relative to the translation group generated by $K$, i.e., $\tilde{C}_{\rm w}(\varrho;K)$. 

Next, as the corrolary of the {\bf Lemma 1}, we have the following relations.\\
{\bf Corollary 3}. Given a state $\varrho$, and a set $\Lambda_{\{k\}}$ of Hermitian operator $K$ with the same nontrivial spectrum $\{k\}$, the maximum normalized TC w-cohrence in $\varrho$ relative to all $K\in\Lambda_{\{k\}}$, is lower bounded as:  
\begin{eqnarray}
\sup_{K\in\Lambda_{\{k\}}}\tilde{C}_{\rm w}(\varrho;K)\ge\sup_{K\in\Lambda_{\{k\}}}\sup_{X\in\Lambda_{\{x\}}}\frac{1}{2}|{\rm Tr}([\tilde{X},\tilde{K}]\varrho)|.
\label{maximal coherence and maximal incompatibility 1}
\end{eqnarray} 
On the other hand, given a fixed $K$, and a collection of states $\varrho$ belonging to a specific set $\Lambda_{\varrho}$, the normalized TC w-coherence maximized over all $\varrho\in\Lambda_{\varrho}$ is lower bounded as 
\begin{eqnarray}
\sup_{\varrho\in\Lambda_{\varrho}}\tilde{C}_{\rm w}(\varrho;K)\ge\sup_{\varrho\in\Lambda_{\varrho}}\sup_{X\in\Lambda_{\{x\}}}\frac{1}{2}|{\rm Tr}(\tilde{K}[\tilde{X},\varrho])|.
\label{maximal coherence and maximal incompatibility 2}
\end{eqnarray} 
Note that in both cases, we must take the supremum over all $X\in\Lambda_{\{x\}}$ whose set of the eigenvectors $\{\ket{x}\}$ span the Hilbert space.  

Finally, we obtain the following uncertainty relation. \\
{\bf Theorem 5}. Given a quantum state $\varrho$, the normalized TC w-coherences relative to the translation group generated by a Hermitian operator $K$ multiplied by that relative to the translation generated by a Hermitian operator $X$ is lower bounded by the average noncommutativity between $\tilde{K}$ and $\tilde{X}$ over $\varrho$ as
\begin{eqnarray}
\tilde{C}_{\rm w}(\varrho;K)\tilde{C}_{\rm w}[\varrho;X]\ge\frac{|{\rm Tr}([\tilde{X},\tilde{K}]\varrho)|^2}{4}.  
\label{uncertainty relation for coherence}
\end{eqnarray}
{\bf Proof}. First, exchanging the role of $K$ and $X$ in Eq. (\ref{coherence and incompatibility 1}) we have 
\begin{eqnarray}
\tilde{C}_{\rm w}[\varrho;X]\ge\sup_{K\in\Lambda_{\{k\}}}\frac{1}{2}|{\rm Tr}([\tilde{X},\tilde{K}]\varrho)|.
\label{coherence and incompatibility 2}
\end{eqnarray}
Multiplying Eqs. (\ref{coherence and incompatibility 1}) and (\ref{coherence and incompatibility 2}), we finally obtain
\begin{eqnarray}
\tilde{C}_{\rm w}(\varrho;K)\tilde{C}_{\rm w}(\varrho;X)&\ge&\sup_{X\in\Lambda_{\{x\}}}\frac{1}{2}|{\rm Tr}([\tilde{X},\tilde{K}]\varrho)|\sup_{K\in\Lambda_{\{k\}}}\frac{1}{2}|{\rm Tr}([\tilde{X},\tilde{K}]\varrho)|\nonumber\\
&\ge&\frac{|{\rm Tr}([\tilde{X}_*,\tilde{K}]\varrho)||{\rm Tr}([\tilde{X},\tilde{K}_*]\varrho)|}{4}\nonumber\\
&\ge&\frac{|{\rm Tr}([\tilde{X},\tilde{K}]\varrho)|^2}{4}, 
\label{uncertainty relation for coherence}
\end{eqnarray}
where $\tilde{X}_*=X_*/|x|_{\rm max}$ and $\tilde{K}_*=K_*/|k|_{\rm max}$, with $X_*\in\Lambda_{\{x\}}$ and $K_*\in\Lambda_{\{k\}}$ are the operators which respectively achieve the maximum in Eqs. (\ref{coherence and incompatibility 1}) and (\ref{coherence and incompatibility 2}). \qed\\
Notice that the above relation takes an analogous form as the KWR uncertainty relation.   

\section{Summary and Remarks\label{Conclusion and Remarks}}  

We showed that the coherence in a quantum state $\varrho$ as an asymmetry relative to the translation group generated by a Hermitian operator $K$, can be quantified in terms of the average absolute imaginary part of the weak value of $K$ with the preselected state $\varrho$, maximized over all possible orthonormal bases of the Hilbert space. We argued that the quantity so defined, i.e., TC w-coherence, satisfies certain properties required for a quantifier of TC coherence or translational asymmetry, in particular, it is nonincreasing under translationally-covariant quantum operations. TC w-coherence can be obtained experimentally via the estimation of the weak value using a number of methods proposed in the literatures, combined with a classical parameter optimization procedure, in the fashion of variational quantum circuit. We showed that TC w-coherence in $\varrho$ relative to the translation group generated $K$ is upper bounded by the standard deviation of $K$ in $\varrho$, and also by the square root of the quantum Fisher information about a parameter contained in a state shifted by the unitary generated by $K$. It is also upper bounded by the maximum total sum of the absolute imaginary part of the associated Kirkwood-Dirac quasiprobability. We obtained a lower bound and derived a relation between the TC w-coherences relative to two generators of translation, having a form analogous to the KWR uncertainty relation. 

That weak value can be used to operationally characterize coherence as asymmetry goes along with the fact that it can also be used to directly reconstruct the quantum state which encodes the coherence and other nonclassical features of quantum systems \cite{Lundeen direct measurement of wave function,Lundeen measurement of KD distribution,Johansen quantum state from successive projective measurement}. It is interesting to remark that the anomalous imaginary part of the weak value \cite{Dressel imaginary weak value and disturbance}, which has received less attention in comparison to its anomalous real part (those that lie outside of the spectrum of the eigenvalues of the observable), offers not only a quantum resource theoretical but also operational framework to study quantum coherence, one of the defining features of quantum systems. Furthermore, by connecting coherence as asymmetry to weak values, we link the nonclassicality encoded in the formal notion of coherence and/or asymmetry, with the nonclassicality captured by the operationally well-defined anomalous complex weak values \cite{Pusey strange weak value and contextuality,Lostaglio TMH quasiprobability fluctuation theorem contextuality,Kunjwal contextuality of non-real weak value}. Finally, it is also interesting to further study the geometrical meaning of the TC w-coherence defined in Eq. (\ref{TC w-coherence}), i.e., to investigate its relation with the other measures of TC coherence based on geometrical characterization of quantum states in Hilbert space and that based on Wigner-Yanase skew information \cite{Wigner-Yanase skew information,Marvian - Spekkens speakable and unspeakable coherence,Luo-Sun coherence from skew information}. 

\begin{acknowledgments}  
This work is partly funded by the Institute for Research and Community Service, Bandung Institute of Technology with the grant number: 2971/IT1.B07.1/TA.00/2021. It is also in part supported by the Indonesia Ministry of Research, Technology, and Higher Education with the grant number: 187/E5/PG.02.00.PT/2022 and 2/E1/KP.PTNBH/2019. The Authors would like to thank the anonymous Referees for constructive criticism and suggestions, and Joel Federicko Sumbowo for useful discussion. 
\end{acknowledgments}

\end{document}